\documentclass[twocolumn,showpacs,preprintnumbers,amsmath,amssymb]{revtex4}

\usepackage{graphicx}
\usepackage{dcolumn}
\usepackage{bm}

\begin{document}

\title{Nonlocal Non-Markovian Effects in Dephasing Environments}
\author{Dong Xie}
\email{xiedong@mail.ustc.edu.cn}
\author{An Min Wang}
 \email{anmwang@ustc.edu.cn}
\affiliation{Department of Modern Physics, University of Science and Technology of China, Hefei, Anhui, China.}

\begin{abstract}
 We study the nonlocal non-Markovian effects through local interactions between two subsystems and the corresponding two environments. It has been found that the initial correlations between two environments can turn a Markovian to a non-Markovian regime with the extra control on the local interaction time. We further research the nonlocal non-Markovian effects from two situations: without extra control, the nonlocal non-Markovian effects only appear under the condition that two local dynamics are non-Markovian-non-Markovian(both of two local dynamics are non-Markovian), or Markovian-non-Markovian, never appear under the condition of Markovian-Markovian; with extra control, the nonlocal non-Markovian effects can occur under the condition of Markovian-Markovian. It shows that the correlations between two environments has an upper bound: only making a flow of information from the environment back to the global system begin finitely earlier than that back to any one of two local systems, not infinitely. Then, due to observing that the classical correlations between two environments have the same function as the quantum correlations, we advise two special ways to distribute classical correlations between two environments without initial correlations. Finally, from numerical solutions in the spin star configuration we obtain that the self-correlation(internal correlation) of each environment promotes the nonlocal non-Markovian effects.
\end{abstract}

\maketitle
\section{Introduction}  
A realistic physical system inevitably interacts with the surrounding environment, leading to lose information from the system to the environment. If the environment can feed back the information to the system in the finite time, it signifies that the non-Markovian effects appear due to the environmental memory. And if the environment only feeds back the information in the infinite time, it means that the whole dynamics is Markovian. The dynamical process of Markovianity can also be treated as a limiting approximation of non-Markovianity\cite{lab1,lab2}.

It is highly interesting to explore the non-Markovian effects, because there are many systems suffering from the strong back-action from environment\cite{lab3}. Hence, the non-Markovianity plays an important role in many respects. Up to now, there are a lot of works about it. Non-Markovianity can assist the formation of steady state entanglement\cite{lab4}; non-Markovian coherent feedback control can also suppress the decoherence\cite{lab5}; and non-Markovian effects are considered in a new theory of polymer reaction kinetics so that the dynamics of polymers can be controlled\cite{lab6}. The direct observation of non-Markovian radiation dynamics has been completed in $3$ dimension bulk photonic crystals\cite{lab7}; and the observation of non-Markovian dynamics of a single quantum dot has also been completed in a micropillar cavity\cite{lab8}.

It is worth to note that the authors, in the Ref.\cite{lab9}, find a new resource for quantum memory effects by the nonlocal non-Markovianity, where they utilize the quantum correlations between two environments and control the local interaction time to turn a Markovian to a non-Markovian regime. It will become more and more interesting in the further research about the nonlocal non-Markovian effects, such as how to enhance and control the nonlocal non-Markovian effects.

In this article, we further discuss the nonlocal non-Markovian effects when the local dynamics are non-Markovian-non-Markovian, Markovian-non-Markovian or Markovian-Markovian. Then, we find that without extra control the nonlocal non-Markovian effects can't appear under the condition of Markovian-Markovian. Besides taking control on the interaction time in the Ref.\cite{lab9}, we find that reducing the strength of interaction also turn a Markovian to a non-Markovian regime. In surprise, increasing the strength of interaction can also do it. Both of two examples in the Ref.\cite{lab9} they consider that the initial correlations of two environments are nonlocal(having quantum correlations). We find that the classical correlations can perform as well as the quantum correlations. In many real situations, the classical correlations between two environments are easier to be formed compared to the quantum correlations. And we advise two special ways to form the classical correlations, leading to the nonlocal non-Markovian effects. Finally, we investigate the non-Markovian effects in the spin star configuration under two different situations: with and without the self-correlation of each environment. We get the numerical solution and obtain that the self-correlation of each environment helps the nonlocal non-Markovian effects.

\section{Theoretical model}

We consider that the initial state of two subsystems is a pure state given by
\begin{equation}
|\Psi_S^{12}(0)\rangle=a|00\rangle+b|01\rangle+c|10\rangle+d|11\rangle.
\end{equation}
Under the local interactions with environments $1$ and $2$, a dephasing map for two qubits of of general form\cite{lab8}
\begin{eqnarray*}
\left(
\begin{array}{cccc}
|a|^2 & ab^*\kappa_2(t) & ac^*\kappa_1(t) & ad^*\kappa_{12}(t)\\
ba^*\kappa_2^*(t) & |b|^2 & bc^*\Lambda_{12}(t) & bd^*\kappa_{1}(t)\\
ca^*\kappa_1^*(t) & cb^*\Lambda^*_{12}(t) & |c|^2 & cd^*\kappa_{2}(t)\\
da^*\kappa_{12}^*(t) & db^*\kappa^*_{1}(t) & dc^*\kappa_2^*(t) & |d|^2\\
\end{array}
\right),
\end{eqnarray*}
Where $\kappa_i(0)=1$ for $i=1$, 2, and 12; $\Lambda_{12}(0)=1$.

The dephasing process of local systems $\rho_S^1(t)$ and $\rho_S^2(t)$ are fully determined by the function $|\kappa_1(t)$ and $|\kappa_2(t)|$. Besides $|\kappa_1(t)$ and $\kappa_2(t)|$, the dephasing process of global systems $\rho_S^1(t)$ also depends on the function $|\kappa_{12}(t)|$ and$|\Lambda^*_{12}(t)|$. So if $|\kappa_1(t)|$ and $|\kappa_2(t)|$ decrease, meanwhile, $|\kappa_{12}(t)|$ or $|\Lambda^*_{12}(t)|$ increases, it is possible that the local systems lose information by the dephasing, but at the same time the global system increases the information. Here we use the measure for non-Markovianity of the dephasing process $\Phi(t)$ \cite{lab2}
\begin{eqnarray}
\mathcal {N}(\Phi)=\textmd{max}_{\rho_{1,2}(0)}\int_{\sigma>0}dt \sigma(t,\rho_{1,2}(0)),
\end{eqnarray}
where $\rho_1$ and $\rho_2$ represent two different states of same system, $\sigma(t,\rho_{1,2}(0)=\frac{d}{dt}D(\rho_1,\rho_2),$ and the trace distance $D(\rho_1,\rho_2)=1/2\textmd{tr}\sqrt{(\rho_1(t)-\rho_2(t))^\dag(\rho_1-\rho_2)}$. $\sigma(t,\rho_{1,2}(0)>0$ represents that the information flows backs to the system from the environment. Namely, the dynamics is non-Markovian. For a two-level system, such as the local system 1, if the state of system recovers coherence at a finite time $t$ ($\frac{d}{dt}|\kappa_1(t)|>0$), the local dynamics is non-Markovian; otherwise, the local dynamics is Markovian. This can be proved as follows. Choose two arbitrary initial states
\begin{eqnarray*}
\left(
\begin{array}{cccc}
a_1 & b_1^*  \\
b_1& d_1 \\
\end{array}
\right) \textmd{and}
\left(
\begin{array}{cccc}
a_2 & b_2^*  \\
b_2 & d_2 \\
\end{array}
\right)
.
\end{eqnarray*}
After the dephasing process, the two states are given by
\begin{eqnarray*}
\left(
\begin{array}{cccc}
a_1 & b_1^*\gamma(t) \\
b_1\gamma^*(t) & d_1 \\
\end{array}
\right) \textmd{and}
\left(
\begin{array}{cccc}
a_2 & b_2^*\gamma(t) \\
b_2\gamma^*(t) & d_2 \\
\end{array}
\right).
\end{eqnarray*}
 Then, get the function $\sigma(t,\rho_{1,2}(0)=\frac{\frac{d}{dt}|\gamma(t)|}{\sqrt{|b_1-b_2|^2|\gamma(t)|^2+(a_1-a_2+d_2-d_1)^2/4}}$.
 So $\sigma(t,\rho_{1,2}(0)>0$ is equivalent to $\frac{d}{dt}|\gamma(t)|>0$.

For a many-level system (global system), as long as the function $\frac{d}{dt}|\kappa_{12}(t)|>0$ or $\frac{d}{dt}|\Lambda_{12}(t)|>0$ at a finite time, the information of some systems must increase,  signifying that the global dynamics is non-Markovian, and vice versa. In other words, in the dephasing environment, whether the global dynamics is non-Markovian depends on $\frac{d}{dt}|\kappa_{12}(t)|>0$ and $\frac{d}{dt}|\Lambda_{12}(t)|>0$(global system) at finite time $t$, besides $\frac{d}{dt}|\kappa_1(t)|>0$ (local system 1) and $\frac{d}{dt}|\kappa_2(t)|>0$ (local system 2).

\section{Three dynamic process}
 We consider that the environments have continuous energy levels $w$ and the corresponding eigenstate $|w\rangle$($\hbar=1$ throughout this article) without loss of generality. The initial state of local system is given by
\begin{eqnarray}
\rho_E^i(0)&=Z_0^i\{\int_0^\infty dw\exp[-(w-w_0^i)^2]|w+c_i\rangle\langle w+c_i|\nonumber\\
&+\int_{-\infty}^0dw\exp[-(w+w_0^i)^2)]|w+c_i\rangle\langle w+c_i|\},
\end{eqnarray}
where for $i=1$ and $2$ (using this notation in the following sections), the energy level $w_0^i\geq0$ and $Z_0^i$ is the normalization coefficient. The energy level $w_0^i$ and $c_i$ can decide whether the local dynamics is
non-Markovian or Markovian.

 Firstly, we discuss that the local dynamics are non-Markovian-non-Markovian. The initial correlations between two environments are the classical correlations. So, without loss of generality, we let the initial state of two environments to be
 \begin{eqnarray}
\rho_E^{12}(0)=&Z_{1}\{\int_0^\infty dw\exp[-(w-1)^2]|w\rangle\langle w|\bigotimes|w\rangle\langle w|\nonumber\\
&+\int_{-\infty}^0dw\exp[-(w+1)^2)]|w\rangle\langle w|\bigotimes|w\rangle\langle w|\}.
\end{eqnarray}
The interaction Hamiltonian is given by
\begin{eqnarray}\label{eqn4}
H_{int}^i=g_i\int_{-\infty}^\infty dw\sigma_z^i\sigma_w^i,
\end{eqnarray}
where $\sigma_w=w|w\rangle\langle w|$; $g_i$ is a coupling constant; and $\sigma_z^i$ is the Pauli operator of system $i$. Then,
\begin{eqnarray}\label{eqn5}
&|\kappa_1(t)|=|\kappa_2(t)|=|2Z_{1}\int_0^\infty dw\exp[-(w-1)^2]\cos(2gwt)|,\nonumber\\
&|\Lambda_{12}(t)|=|2Z_{1}\int_0^\infty dw\exp[-(w-1)^2]\cos(4gwt)|,\nonumber\\
&|\kappa_{12}(t)|=1,
\end{eqnarray}
where $g=g_i$ like $g$ in the following Eq.(\ref{eqn7}) and Eq.(\ref{eqn9}).
\begin{figure}[h]
\includegraphics[scale=1]{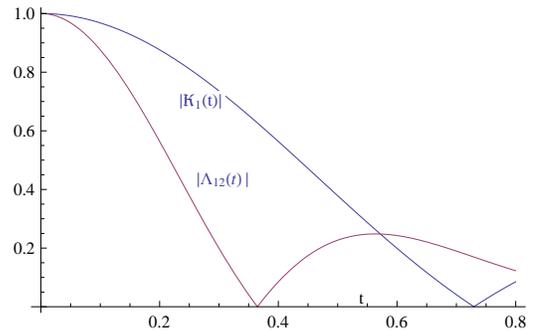}
 \caption{\label{fig.1}From the Eq.(\ref{eqn5}), $|\kappa_1|$ changes with time $t$, in comparison with $|\Lambda_{12}|$. Here, the coupling constant $g=1$.   }
 \end{figure}
 As shown in Fig.1, at time $t=0.36$, $|\Lambda_{12}|$ begins to increase when $\kappa_1$ and $\kappa_2$ keeps on decreasing. It means that the flow of information from the environment back to the global system begins earlier than both of two local subsystems.

 Secondly, we consider that the local dynamics are Markovian-non-Markovian. The initial state of two environments is given by
  \begin{eqnarray}
\rho_E^{12}(0)=&Z_{1}\{\int_0^\infty dw\exp[-w^2]|w\rangle\langle w|\bigotimes|w+1\rangle\langle w+1|\nonumber\\
&+\int_{-\infty}^0dw\exp[-w^2)]|w\rangle\langle w|\bigotimes|w+1\rangle\langle w+1|\}.
\end{eqnarray}
The interaction Hamiltonian is same as Eq.(\ref{eqn4}). Then we obtain
\begin{eqnarray}\label{eqn7}
&|\kappa_1(t)|=|2Z_{1}\int_0^\infty dw\exp[-w^2]\cos(2gwt)|,\nonumber\\
&|\kappa_2(t)|=|2Z_{1}\int_0^\infty dw\exp[-w^2]\cos[g(2w+2)t]|,\nonumber\\
&|\Lambda_{12}(t)|=|2Z_{1}\int_0^\infty dw\exp[-w^2]\cos[g(4w+2)t]|,\nonumber\\
&|\kappa_{12}(t)|=|2Z_{1}\int_0^\infty dw\exp[-w^2]\cos(2gt)|.
\end{eqnarray}
\begin{figure}[h]
\includegraphics[scale=1.2]{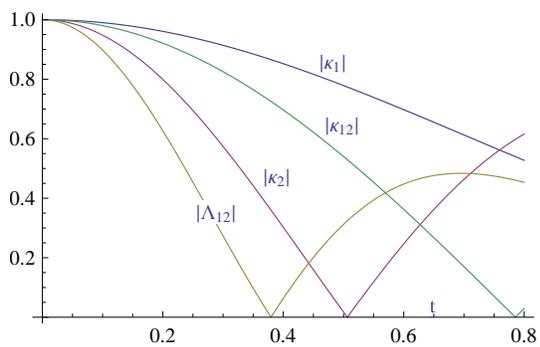}
 \caption{\label{fig.2}From the Eq.{\ref{eqn7}}, $|\kappa_1|$, $|\kappa_2|$, $|\kappa_{12}|$, and $|\Lambda_{12}|$ change with time $t$.  Here, the coupling constant $g=1$.   }
 \end{figure}
 In Fig.2, we can also see that after a while, $\Lambda_{12}$ begins to increase when $\kappa_1$ and $\kappa_2$ keeps on decreasing, meaning that the flow of information from the environment back to the global system begins earlier than that back to  the local subsystem $2$. Where, $|\kappa_1|$ always reduces($\frac{d}{dt}|\kappa_1|<0$ for all time), representing that the dynamics of subsystem $1$  is Markovian.

 Finally, we consider that the local dynamics are Markovian-Markovian. A classical initial state of two environments is described by
 \begin{eqnarray}\label{eqn8}
\rho_E^{12}(0)=&Z_{1}\{\int_0^\infty dw\exp[-w^2]|w\rangle\langle w|\bigotimes|w\rangle\langle w|\nonumber\\
&+\int_{-\infty}^0dw\exp[-w^2]|w\rangle\langle w|\bigotimes|w\rangle\langle w|\}.
\end{eqnarray}
 Furthermore, use the same interaction Hamiltonian (see the Eq.\ref{eqn4}) to get
 \begin{eqnarray}\label{eqn9}
&|\kappa_1(t)|=|\kappa_2(t)|=|2Z_{1}\int_0^\infty dw\exp[-w^2]\cos(2gwt)|,\nonumber\\
&|\Lambda_{12}(t)|=|2Z_{1}\int_0^\infty dw\exp[-w^2]\cos[4gwt]|,\nonumber\\
&|\kappa_{12}(t)|=1.
\end{eqnarray}
\begin{figure}[h]
\includegraphics[scale=1.2]{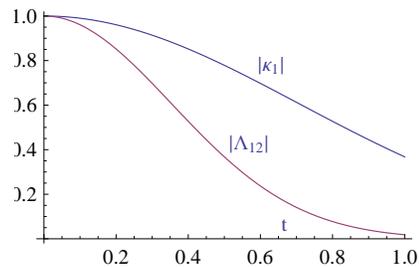}
 \caption{\label{fig.3}From the Eq.(\ref{eqn9}), $|\kappa_1|$ and $|\Lambda_{12}|$ change with time $t$.  Here, the coupling constant $g=1$.   }
  \end{figure}
  From Fig.3, we know that $\Lambda_{12}$ always decreases, never increases. When the local dynamics are Markovian-Markovian, we can't find a initial state of two environments to make the nonlocal non-Markovian effects appear, even the existence of quantum correlations between two environments.

 From the above three dynamic process and corresponding figures, we can get a conclusion that the correlations between two environments only make the flow of information from the environment back to the global system begins finitely earlier than that back to  any one of two local subsystems. In another words, there is a upper bound on the function of the correlations between two environments. So if the local dynamics are Markovian-Markovian, the correlations can't shorten the infinite start time of the flow of information from the environment back to the global system to a finite one. Namely, the global dynamics is still Markovian when the local dynamics are Markovian-Markovian.
\section{Extra control}
 However, all foregoing analysis bases on without extra control(representing that the whole Hamiltonian of systems and two environments is independent of time). So if there are other control, it is possible to obtain the nonlocal non-Markovian effects under the condition that the local dynamics are Markovian-Markovian.

 In the Ref.\cite{lab9}, they control the time of two local interactions to turn a Markovian to a non-Markovian regime. Here, we find that reducing the coupling strength $g_i$(here, denotes that reducing the rate of local dephasing) can do it. In amazement, increasing the coupling strength (increase the rate of local dephasing) can also perform as well.

 Let the initial state of two environments to be the Eq.(\ref{eqn8}), and the interaction Hamiltonian is given by Eq.(\ref{eqn4}). So the local dynamics are Markovian-Markovian. Without extra control, the global dynamics also is Markovian.

 Firstly, we consider reducing the coupling strength. Initially, the coupling strength $g_1=3$ and $g_2=2$; when $t=1$, reduce $g_1$ from $3$ to $1$. Then, we get
 \begin{eqnarray}\label{eqn10}
 &|\kappa_1(t')|=|2Z_{1}\int_0^\infty dw\exp[-w^2]\cos(wt'+3w)|,\nonumber\\
 &|\kappa_2(t')|=|2Z_{1}\int_0^\infty dw\exp[-w^2]\cos[2wt'+2w]|,\nonumber\\
 &|\Lambda_{12}(t')|=|2Z_{1}\int_0^\infty dw\exp[-w^2]\cos[3wt'+5w]|,\nonumber\\
 &|\kappa_{12}(t')|=|2Z_{1}\int_0^\infty dw\exp[-w^2]\cos(w-wt')|,
\end{eqnarray}
where $t'=t-1$.
 \begin{figure}[h]
\includegraphics[scale=1.2]{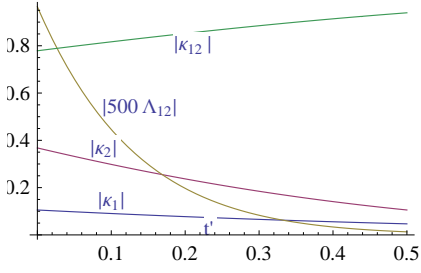}
 \caption{\label{fig.4}From the Eq.(\ref{eqn10}), $|\kappa_1|$, 500$|\Lambda_{12}|$, $|\kappa_2|$ and $|\kappa_{12}|$change with time $t'$. Where, in order to better compare them, we replace the $|\Lambda_{12}|$ by the $500|\Lambda_{12}|$.    }
  \end{figure}
 As shown by Fig.4, it is obviously that $|\kappa_{12}|$ increases, signifying that reducing the strength of coupling $g_1$ can turn the Markovian to the non-Markovian regime.

 Next, we discuss about increasing the coupling strength. Initially, the coupling strength $g_1=2$ and $g_2=1$; when $t=1$, increase $g_2$ from $1$ to $3$. Then, we can obtain
\begin{eqnarray}\label{eqn11}
 &|\kappa_1(t')|=|2Z_{1}\int_0^\infty dw\exp[-w^2]\cos(wt'+2w)|,\nonumber\\
 &|\kappa_2(t')|=|2Z_{1}\int_0^\infty dw\exp[-w^2]\cos[2wt'+w]|,\nonumber\\
 &|\Lambda_{12}(t')|=|2Z_{1}\int_0^\infty dw\exp[-w^2]\cos[5wt'+3w]|,\nonumber\\
 &|\kappa_{12}(t')|=|2Z_{1}\int_0^\infty dw\exp[-w^2]\cos(w-wt')|,
\end{eqnarray}
where $t'=t-1$.
\begin{figure}[h]
\includegraphics[scale=1.2]{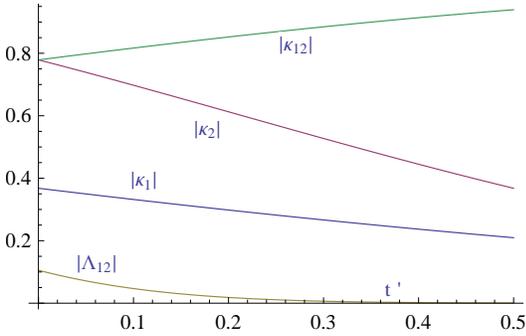}
 \caption{\label{fig.4}From the Eq.(\ref{eqn11}), $|\kappa_1|$, $|\Lambda_{12}|$, $|\kappa_2|$ and $|\kappa_{12}|$change with time $t'$.     }
  \end{figure}
  From Fig.5, $|\kappa_{12}|$ also increases, representing that the nonlocal non-Markovian effects appears when the coupling strength increases.

\section{Forming classical correlations}
 Two nonlocal environments often haven't any correlations. It is very hard to create the quantum correlations between two macroscopical environments. But the classical correlations can be formed relatively easy. And, as shown by the above sections, the classical correlations can perform very well for the nonlocal non-Markovian effects. Then, we advise two ways to form the classical correlations between two environments.

Firstly, we consider that a Bell state $1/\sqrt{2}(|00\rangle+|11\rangle)$ disentangles by the local interaction with two environments, where the environments are composed of bosons. Then let the system locally interact with the environments.

The interaction Hamiltonian is given by
\begin{eqnarray}
 H_{int}^i(t)&=\chi_i(t)\sum_{k_i=1}^{n_i}g_{k_i}\sigma_z^i(b_{k_i}^\dagger+b_{k_i})\nonumber\\
 &+{\chi}'_i(t)\sum_{k_i=1}^{n_i}g_{k_i}{\sigma '}_z^{i}(b_{k_i}^\dagger+b_{k_i}),
\end{eqnarray}
where the first term on the right hand side is the interaction Hamiltonian between the system and the environments; the second term is the interaction Hamiltonian between the entanglement system(another system which is initially in the Bell state and $\sigma_Z '^i$ is the Pauli operator of corresponding subsystem $i$) and the same environments; the function ${\chi}'_i(t)=1$ for $0\leq t\leq t'$ and zero otherwise; ${\chi}_i(t)=1$ for $t_i^s\leq t\leq t_i^f$ and zero otherwise. Here, $t_i^s$ and $t_i^f$ denote the time when the interaction is switched on and switched off in system $i$, respectively; 0 and $t'$ denote the time in the entanglement system. We denote that $t_i(t)=\int_0^t\chi_i(t')dt'$ and  $t'_i(t)=\int_0^t{\chi}'_i(t')dt'$.
The Hamiltonian of environments is described by
\begin{eqnarray}
 H_E^{12}&=\sum_{k_1=1}^{n_1}w_{k_1}b^\dagger_{k_1} b_{k_1}+\sum_{k_2=1}^{n_2}w_{k_2}b^\dagger_{k_2} b_{k_2}.
\end{eqnarray}
The initial state of environments is in the thermal equilibrium state $\rho_E^{12}=1/Z_{12}\exp(-\beta H_E^{12})$, where $Z_{12}$ is the partition function.

Then, using weak coupling approximation $[\exp(\alpha_k b^\dag-\alpha_k^* b), \exp(-i w_k b_k^\dag b_k)]\approx0$, and performing continuum limit with Ohmic spectrum density $A_iw\exp(-w/\Omega_i)$\cite{lab10,lab11,lab12} ($A_i$ is coupling constant and $\Omega_i$ is frequency cutoff), we get
\begin{eqnarray}\label{eqn14}
&|\kappa_{1}|=|\exp(-\Gamma_1)\cos[\int dwA_1\exp(-w/\Omega_1)\xi_1]|,\nonumber\\
&|\Lambda_{12}|=|\exp(-\Gamma_1-\Gamma_2)\cos[\sum_{i=1}^2\int dwA_i\exp(-w/\Omega_i)\xi_i]|,\nonumber\\
&\textmd{in which},\nonumber\\
& \xi_i=2\sin [w(t_i(t)-t'_i(t))]-2\sin [w t_i(t)]+2\sin [w t'_i(t)], \nonumber\\
&\Gamma_i=\int dwA_i\exp(-w/\Omega_i)\coth(2w/\beta)(1-\cos[w t_i(t)],\nonumber\\
&\alpha_k=g_k\frac{1-\exp(iw_kt)}{w_k}.
\end{eqnarray}
As shown in Fig.6, the information begins to flow back to the global system when $|\Lambda_{12}|$ begins to increase, meaning that the correlations between the two environments is formed to make the nonlocal non-Markovian effects emerge. And we find the Bell state have the same function as the classical state $1/2|00\rangle\langle00|+1/2|11\rangle\langle11|$ on forming the classical correlations of two environments, so it has the robustness against some noise. If one want to form stronger correlations, it just need to increase the number of Bell state.
\begin{figure}
\includegraphics[scale=1.2]{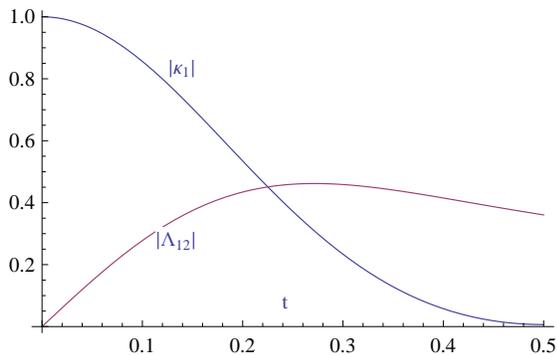}
 \caption{\label{fig.6}From the Eq.(\ref{eqn14}), $|\kappa_1|$ and $|\Lambda_{12}|$ change with time $t$. Here, the parameter $\Gamma_2=0.5$, $A_1=1$, $\beta=0.2$, $\xi_2=\pi/2$, $t'_1(t)=t'_2(t)=1$ and $t_1(t)=t$. }
  \end{figure}

  Then, the second way is that after two subsystems interact locally with the corresponding two environments without the initial correlations for some time, exchanging the locations of two subsystems generate that they interact locally with each other initial environment. Similar calculation as the above way which is immune to data, it is easy to observe the nonlocal non-Markovian effects. The reason is that the system loses information, leading to the correlations between two environments. And it is necessary to exchange the locations of two subsystems because if don't exchange the locations, the  correlations between the system and the two environments will keep the system losing information, never lead to the appearance of nonlocal non-Markovian effects. The benefit of this way is that it don't need other correlational states to form the correlations between two environments. Obviously, it can't form very strong correlations like the first way. Furthermore, it can't create the strong non-Markovian effects.
 \section{The self-correlation promoting the nonlocal non-Markovian effects}
  In this section, we explore whether the self-correlation of two environments promotes the nonlocal non-Markovian effects. We consider a simple model: the spin star configuration\cite{lab13}.

  The interaction Hamiltonian is described by
\begin{eqnarray}
  &H_{int}=\sum_{i=1}^2\sum_{j=1}^{n_i}\eta_i(t)g_{ij}\sigma_{ij}^{z}\sigma_{iS}^z,
\end{eqnarray}
where $\eta_i(t)=1$ for $ t_i^s\leq t\leq t_i^f$ and zero otherwise; $\sigma_{ij}^{z}$ and $\sigma_{iS}^z$ are the Pauli operators of environment $i$ and subsystem $i$ respectively.
The initial two environments is in the thermal equilibrium state $\rho_E^{12}=1/Z_{12}\exp[-\beta H_E^{12}]$, in which,
the Hamiltonian of two environments
  $H_E^{12}=\sum_{i=1}^2B_iS_i^z+\alpha S_1^zS_2^z$.
The operator $S_i^z=\sum_{j=1}^{n_i}1/2\sigma_{ij}^z$ for having not the self-correlation of each environment; and $S_i^z=\sum_{j=1}^{n_i}(1/2\sigma_{ij}^z+J_i/B_i\sum_{<mn>}\sigma_{im}^z\sigma_{in}^z)$ for having the self-correlation of each environment.
Then we obtain
\begin{eqnarray}\label{eqn16}
  &|\kappa_1(t)|=|Tr[\exp[-2i\sum_{j=1}^{n_1}\int_0^tdt'\eta_1(t')g_{1j}\sigma_{1j}^z]\rho_E^{12}]|,\nonumber\\
  &|\Lambda_{12}(t)|=|Tr[\exp[-2i\sum_{i=1}^2\sum_{j=1}^{n_i}\int_0^tdt'\eta_i(t')g_{ij}\sigma_{ij}^z]\rho_E^{12}]|.
\end{eqnarray}
\begin{figure}
\includegraphics[scale=1.2]{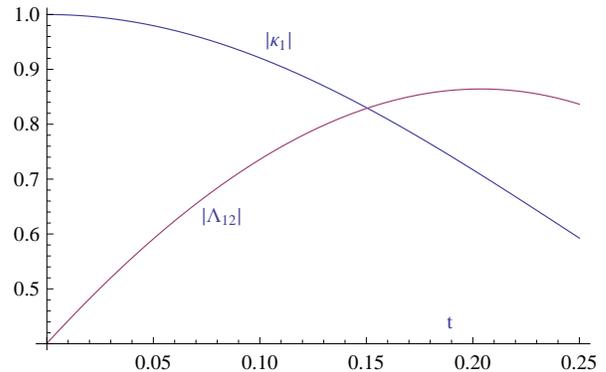}
 \caption{\label{fig.7}$|\kappa_1|$ and $|\Lambda_{12}|$ change with time $t$ for having the self-correlation. Here, the parameter: $n_i=5$, $\alpha=4$, $\beta=0.01$, $B_i=2$, $J_i=10$, $\int_0^tdt'\eta_2(t')=0.2$, $\int_0^tdt'\eta_1(t')=t$ and $g_{ij}=1$ for $i=1,2$; $j=1,...,5$.}
  \end{figure}
  \begin{figure}
\includegraphics[scale=1.2]{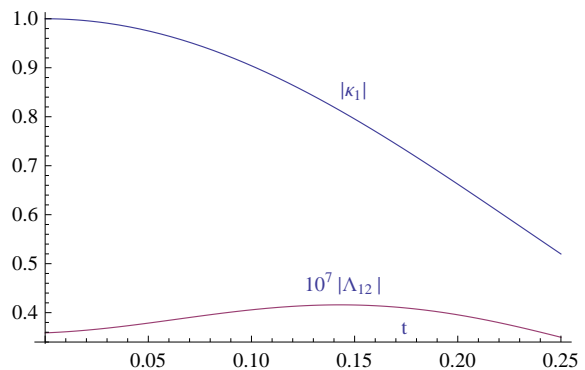}
 \caption{\label{fig.8}$|\kappa_1|$ and $10^7|\Lambda_{12}|$ change with time $t$ without the self-correlation. Here, the parameter: $n_i=5$, $\alpha=4$, $\beta=0.01$, $B_i=2$, $\int_0^tdt'\eta_2(t')=0.785$, $\int_0^tdt'\eta_1(t')=t$ and $g_{ij}=1$ for $i =1,2$; $j=1,...,5$.}
  \end{figure}
  Comparing Fig.7 with Fig.8 which are the numerical graph of Eq.(\ref{eqn16}), it is obviously that the nonlocal non-Markovian effects with the help of self-correlation is stronger, according to the growth of $\Lambda_{12}(t)$.
  \section{Conclusion}
  In this article we explore that a quantum system composed of two subsystems locally interacts with two environments in detail. We obtain that the function of correlations of two environments has an upper bound: only make a flow of information from the environment back to the global system start finitely earlier than that back to any one of two local systems, not finitely(when the whole Hamiltonian of system and two environments is dependent of time). It means that nonlocal non-Markovian effects can't appear when both of two local dynamics are Markovian. Of course, extra control can turn a Markovian to a non-Markovian regime. Besides the control of local interaction time, reducing the coupling strength of local interaction(representing that reduce the rate of losing information of subsystems) can make the nonlocal non-Markovian effects appear. Surprisingly, enhancing the the couping strength of local interaction( increase the rate of losing information of subsystems) can also make it. And we advise two ways  which is easy in the experiment to form the classical correlations between two environments without initial correlations. Finally, we obtain that the self-correlation of two environments can promote the nonlocal non-Markovian effects.

 Recently, the Ref.\cite{lab14} accessed the non-Markovianity based on ideas of divisibilities of channels; the Ref.\cite{lab15} proposed a way to quantify the memory effects in the spin-boson model; the Ref.\cite{lab16} advised two different approaches to quantify non-Markovianity; and the Ref.\cite{lab17} quantified non-Markovianity via correlations. It is also very interesting to quantify the nonlocal non-Markovianity in different ways, for a better understanding and characterization of nonlocal non-Markovian effects in more complex systems. The nonlocal non-Markovian effects means the nonlocal memory of global environments, which may be used to perform some quantum information process, such as quantum memory\cite{lab18} and quantum error correction\cite{lab19}. And this article will be meaningful for how to utilize and control the nonlocal non-Markovian effects by controlling the local dynamics in the future works.
\section{Acknowledgments}
This work was supported by the National Natural Science Foundation of China under Grant No. 10975125.


\begin{thebibliography}{0}

\bibitem{lab1} Herbert Spohn 1980 \textit{Rev. Mod. Phys.} {\bf52} 569.
\bibitem{lab2}Heinz-Peter Breuer, Elsi-Mari Laine, and Jyrki Piilo 2009 \textit{Phys. Rev. Lett.} {\bf103} 210401.
\bibitem{lab3}A. Ishizaki and G. R. Fleming 2009 \textit{J. Chem. Phys.} {\bf130} 234110; 2009 {\bf130} 234111;
    B. Bellomo, R. Lo Franco, and G. Compagno 2007 \textit{Phys. Rev. Lett}. {\bf99} 160502; Fleming, and K. B. Whaley, 2010 \textit{Nature Phys.} {\bf6} 462; A. G. Dijkstra and Y. Tanimura 2010 \textit{Phys. Rev. Lett.} {\bf104} 250401; J.-Q. Liao, J.-F. Huang, L.-M. Kuang, and C. P. Sun 2010 \textit{Phys. Rev.} A {\bf82} 052109; A. Imamoglu 1994 \textit{Phys. Rev. A} {\bf50} 3650; Wang Xiao-Yun, Ding Bang-Fu, and Zhao He-Ping 2013 \textit{Chin. Phys. B} {\bf22} 040308; Ding Bang-Fu, Wang Xiao-Yun, Tang Yan-Fang, Mi Xian-Wu, and Zhao He-Ping 2011 \textit{Chin. Phys. B} {\bf20} 060304.
\bibitem{lab4}Susana F. Huelga, $\acute{A}$ngel Rivas, and Martin B. Plenio 2012 \textit{Phys. Rev. Lett.} {\bf108} 160402.
\bibitem{lab5} Shi-Bei Xue, Re-Bing Wu, Wei-Min Zhang, Jing Zhang, Chun-Wen Li, and Tzyh-Jong Tarn 2012 \textit{Phys. Rev.} A {\bf86} 052304.
\bibitem{lab6}T. Gu$\acute{e}$rin, O. B$\acute{e}$nichou, and R. Voituriez 2012 \textit{Nature Chemistry} {\bf4} 568-573.
\bibitem{lab7}Ulrich Hoeppe, Christian Wolff, Jens Kuchenmeister, Jens Niegemann, Malte Drescher, Hartmut Benner, and Kurt Busch 2012 \textit{Phys. Rev. Lett.} {\bf108} 043603.
\bibitem{lab8}K. H. Madsen, S. Ates, T. Lund-Hansen, A. L$\ddot{o}$ffler, S. Reitzenstein, A. Forchel, and P. Lodahl 2011 \textit{Phys. Rev. Lett.} {\bf106} 233601.
\bibitem{lab9}Elsi-Mari Laine, Heinz-Peter Breuer, Jyrki Piilo, Chuan-Feng Li, and Guang-Can Guo 2012 \textit{Phys. Rev. Lett.} {\bf108} 210402.
\bibitem{lab10}H.-P. Breuer and F. Petruccione 2002 \textit{The Theory of Open Quantum Systems} (Oxford University Press, New York).
\bibitem{lab11}A. J. Leggett, S. Chakravarty, A. Dorsey, M. Fisher, A. Garg, and W. Zwerger 1987 \textit{Rev. Mod. Phys.} {\bf59} 1.
\bibitem{lab12}U. Weiss 2008 \textit{Quantum Dissipative Systems} (World Scientific, Singapore).
\bibitem{lab13} Heinz-Peter Breuer, Daniel Burgarth, and Francesco Petruccione 2004 Phys. Rev. B {\bf70} 045323.
\bibitem{lab14}M. M. Wolf, J. Eisert, T. S. Cubitt, and J. I. Cirac 2008 \textit{Phys. Rev. Lett.} {\bf101} 150402.
\bibitem{lab15}Govinda Clos and Heinz-Peter Breuer 2012 \textit{Phys. Rev.} A {\bf86} 012115.
\bibitem{lab16}$\acute{A}$ngel Rivas, Susana F. Huelga, and Martin B. Plenio 2010 \textit{Phys. Rev. Lett.} {\bf105} 050403.
\bibitem{lab17}Shunlong Luo, Shuangshuang Fu, and Hongting Song 2012 \textit{Phys. Rev.} A {\bf86} 044101.
\bibitem{lab18}Nicolas Sangouard, Christoph Simon, Hugues de Riedmatten, and Nicolas Gisin 2011 \textit{Rev. Mod. Phys.} {\bf83} 33.
\bibitem{lab19}Gerardo A. Paz-Silva, A. T. Rezakhani, Jason M. Dominy, and D. A. Lidar 2012 \textit{Phys. Rev. Lett.} {\bf108} 080501.

\end{thebibliography}
 \end{document}